\documentclass{article}

\usepackage[english]{babel}
\usepackage[T1]{fontenc}
\usepackage[utf8]{inputenc}

\usepackage{amsmath,amssymb}
\usepackage{spconf}
\usepackage{xcolor}
\usepackage{booktabs}
\usepackage{graphicx}
\usepackage{multirow}
\usepackage{tabularx}

\usepackage{pgf}
\usepackage{pgffor}
\usepgflibrary{plothandlers}

\graphicspath{{images/}}

\title{Reconstruction of Images Taken by a Pair of Non-Regular Sampling Sensors Using Correlation Based Matching}
\name{Markus Jonscher, Jürgen Seiler, Thomas Richter, Michel Bätz, and André Kaup \thanks{This work has been supported by the Deutsche Forschungsgemeinschaft
(DFG) under contract number KA 926/5-1.}}
\address{Multimedia Communications and Signal Processing \\ Friedrich-Alexander University Erlangen-Nürnberg, Cauerstr. 7, 91058 Erlangen, Germany}

\begin{document}
\ninept
\maketitle

\begin{abstract}
Multi-view image acquisition systems with two or more cameras can be rather costly due to the number of high resolution image sensors that are required. Recently, it has been shown that by covering a low resolution sensor with a non-regular sampling mask and by using an efficient algorithm for image reconstruction, a high resolution image can be obtained. 
In this paper, a stereo image reconstruction setup for multi-view scenarios is proposed. A scene is captured by a pair of non-regular sampling sensors and by incorporating information from the adjacent view, the reconstruction quality can be increased. Compared to a state-of-the-art single-view reconstruction algorithm, this leads to a visually noticeable average gain in PSNR of $0.74$ dB.
\end{abstract}


\vspace*{-0.1cm}
\section{Introduction}
\label{sec:intro}
\vspace*{-0.2cm}
There are many applications in image processing requiring acquisition systems that capture a scene by two or more cameras. In 3D reconstruction~\cite{Seitz2006}, a 3D model can be generated from a set of digital images taken by multiple cameras. Surveillance applications~\cite{Dallmeier2012} also tend to use multi-focal sensor systems for video surveillance of wide areas. In entertainment systems, a pair of cameras is used for augmented reality and 3D image acquisition. Recently, systems with three image sensors~\cite{Gere2013} have been considered for mobile devices. They are aiming at capturing photographs with better quality and may also introduce other features like virtual keyboards. In all of these applications, the correlation between the captured images is very high since they describe almost the same scene. One possibility of taking the similarity between such multi-view images into account is the estimation of the disparity between them. The disparity results from the displacement of objects between these images. By employing the disparity, imaging flaws like dead pixels and coloring or lightning problems in one sensor can be replaced with valid image data from other sensors. The acquisition of high quality multi-view image data, however, can be rather costly since multiple sensors with high resolution are required. Additionally, this leads to higher power consumption since more pixels have to be read out by the sensor array and further components like storage, compression and image processing get more complex.

Recently, it has been shown in~\cite{Schoeberl2011a} that it is possible to obtain a high resolution (HR) image by covering a low resolution (LR) image sensor with a non-regular sampling mask. Using an efficient image reconstruction algorithm gives an HR image which has been originally taken by an LR sensor. This allows for replacing HR sensors by LR sensors while retaining almost the same visual quality.
\begin{figure}
	\centering
	\def\svgwidth{\columnwidth}	
	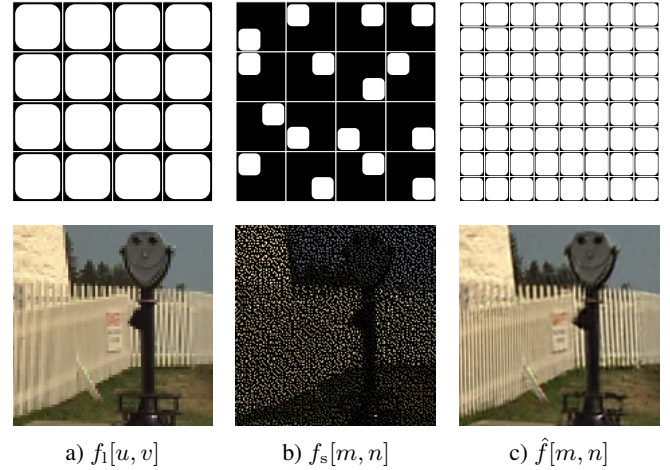	
	\caption{Top row: A low resolution sensor (a) is covered with a non-regular sampling mask (b) and by using an efficient reconstruction algorithm an image with high resolution can be obtained (c). Bottom row: Corresponding image examples.}
	\label{fig:reconstruction_pipeline}
	\vspace*{-0.3cm}
\end{figure}
In Fig.~\ref{fig:reconstruction_pipeline}, the principle of this approach is shown. The first row contains an LR sensor with large pixels yielding an LR image $f_{\mathrm{l}}[u,v]$. The white area of the sensor is sensitive to light and corresponding image detail examples are visualized in the second row. An LR sensor that is covered by a non-regular sampling mask gives an HR image $f_{\mathrm{s}}[m,n]$, where due to the masking, pixel information is lost. $(m,n)$ depict the spatial coordinates on the HR grid and $(u,v)$ on the LR grid. The masking is done in a way that each large pixel of an LR sensor is divided into four quadrants, where three of them are randomly covered. As a consequence, just $1/4$ of the large pixel is sensitive to light anymore. For that reason, this is called $1/4$~sampling. Missing pixels of $f_{\mathrm{s}}[m,n]$ have to be reconstructed by a suitable reconstruction method to obtain an HR image $\hat{f}[m,n]$. By applying Frequency Selective Extrapolation (FSE)~\cite{Seiler2010} for image reconstruction, a high image quality can be achieved. Compared to an image taken by an LR sensor, twice the resolution in both spatial dimensions is possible when using $1/4$~sampling. Masking of an LR sensor, however, is not restricted to $1/4$ sampling. Other potential masks may employ $1/9$ or $1/16$~sampling.

In this paper, an image reconstruction setup that is intended to be used in multi-view scenarios with several cameras is proposed. 
To reduce costs in such applications, the idea of non-regular sampling can be applied and by using dependencies between the camera sensors, the reconstruction quality can be further enhanced.
Without loss of generality, a scene is taken by a pair of non-regular sampling sensors with low resolution to avoid the need for two HR sensors.
Since the image reconstruction quality of FSE highly depends on the number of available sampling points, valid pixel information from one image can be projected into the missing area of the other image. This way, new sampling points are defined and can be used as additional support for the final image reconstruction. For warping pixels from one image into the other image, disparity maps are required. Since typically no such maps are directly available, they can be computed with the help of correlation based matching (CBM). Simulation results show that in case of multi-view scenarios with several cameras using non-regular sampling, it is more beneficial to combine these sensors than process each sensor separately. 

The paper is organized as follows: The next section covers the basic principle of Frequency Selective Extrapolation for image reconstruction and Section \ref{sec:proposed_setup} presents the proposed stereo reconstruction setup. Simulation results are given in Section \ref{sec:results} and Section \ref{sec:conclusion} concludes this paper.

\vspace*{-0.1cm}
\section{Reconstruction of non-regular sampled image data using FSE}
\label{sec:fse}
\vspace*{-0.2cm}
\begin{figure}
	\centering
	\def\svgwidth{\columnwidth}	
	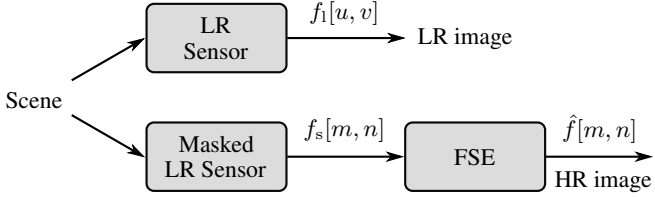
	\caption{Different possibilities for image acquisition using an LR sensor. Upper branch: A scene is taken by an LR sensor leading to an LR image $f_{\mathrm{l}}[u,v]$. Lower branch: The same scene is taken by a masked LR sensor leading to an incomplete HR image $f_{\mathrm{s}}[m,n]$ which is then reconstructed using FSE.}
	\label{fig:block_diagram}
\end{figure}
The proposed stereo reconstruction scenario is based on the idea of using non-regular sampling sensors with low resolution in order to obtain an HR image. Due to partially covering these LR sensors, lost pixels on the resulting HR grid have to be reconstructed. 
FSE iteratively generates the sparse signal model
\begin{equation}
	g[m,n] = \sum_{(k,l)\in\mathcal{K}}\hat{c}_{(k,l)}\varphi_{(k,l)}[m,n]
\end{equation}
as a superposition of Fourier basis functions $\varphi_{(k,l)}[m,n]$ weighted by the expansion coefficients $\hat{c}_{(k,l)}$. The indices of all basis functions used for model generation are summarized in set $\mathcal{K}$. In every iteration, one basis function is selected, its corresponding weight estimated and then added to the model. Missing pixels in the HR image are then replaced using the model $g[m,n]$. 

Fig.~\ref{fig:block_diagram} illustrates the principle of employing non-regular sampling and FSE for image reconstruction. The upper branch shows how a scene is taken by a regular camera with an LR sensor giving an LR image $f_{\mathrm{l}}[u,v]$. In the lower branch, the same scene is taken by the same LR sensor which is partially covered by a non-regular sampling mask. This leads to an HR image $f_{\mathrm{s}}[m,n]$, where due to the masking, pixel information is lost. The reconstruction of this image is done by FSE resulting in a full HR image $\hat{f}[m,n]$. In this paper, FSE additionally uses a frequency weighting function~\cite{Schoeberl2011a} and an optimized processing order~\cite{Seiler2011} as extensions. An extensive discussion on the FSE parameters can be found in these contributions.

\renewcommand{\arraystretch}{1.1}
\setlength{\tabcolsep}{3mm}
\begin{table*}[!t]	
	\caption{PSNR results in dB and corresponding SSIM values for several images from the 2006 Middlebury stereo dataset. All values are taken for the left image sensor and for a sensor distance of 160 mm.}
	\label{tab:psnr_results}
	\vspace*{0.2cm}
	\centering	
	\begin{tabularx}{\textwidth}{lp{0.3cm}cccp{0.4cm}ccc}
		\toprule
		                                                                                       &        & $\text{FSE-SV}_{\text{(PSNR)}}$ & $\text{proposed}_{\text{(PSNR)}}$ & $\Delta_{\text{(PSNR)}}$ &          & $\text{FSE-SV}_{\text{(SSIM)}}$ & $\text{proposed}_{\text{(SSIM)}}$ & $\Delta_{\text{(SSIM)}}$ \\ \midrule
		\it{Aloe}                                                                              &        &             $34.50$             &         $\mathbf{35.93}$          &          $1.43$          &          &            $0.9924$             &         $\mathbf{0.9952}$         &         $0.0028$         \\
		\it{Cloth1}                                                                            &        &             $38.47$             &         $\mathbf{39.81}$          &          $1.34$          &          &            $0.9955$             &         $\mathbf{0.9972}$         &         $0.0017$         \\
		\it{Flowerpots}                                                                            &        &             $47.27$             &         $\mathbf{47.30}$          &          $0.03$          &          &            $0.9986$             &         $\mathbf{0.9987}$         &         $0.0001$         \\
		\it{Midd2}                                                                             &        &             $35.16$             &         $\mathbf{35.79}$          &          $0.63$          &          &            $0.9911$             &         $\mathbf{0.9921}$         &         $0.0010$         \\
		\it{Monopoly}                                                                             &        &             $33.79$             &         $\mathbf{34.46}$          &          $0.67$          &          &            $0.9920$             &         $\mathbf{0.9935}$         &         $0.0015$         \\
		\it{Plastic}                                                                           &        &        $\mathbf{47.43}$         &              $47.19$              &  $\!\!\!\!\!\:\!-0.24$   &          &            $0.9990$             &             $0.9990$              &         $0.0000$         \\ 
		\it{Rocks2}                                                                            &        &             $42.07$             &         $\mathbf{43.10}$          &          $1.03$          &          &            $0.9960$             &         $\mathbf{0.9973}$         &         $0.0013$         \\ \bottomrule
	\end{tabularx}
	\vspace*{-0.3cm}
\end{table*}

\vspace*{-0.1cm}
\section{Proposed Stereo Reconstruction Setup}
\label{sec:proposed_setup}
\vspace*{-0.2cm}
\begin{figure}[!t]
	\centering
	\def\svgwidth{\columnwidth}	
	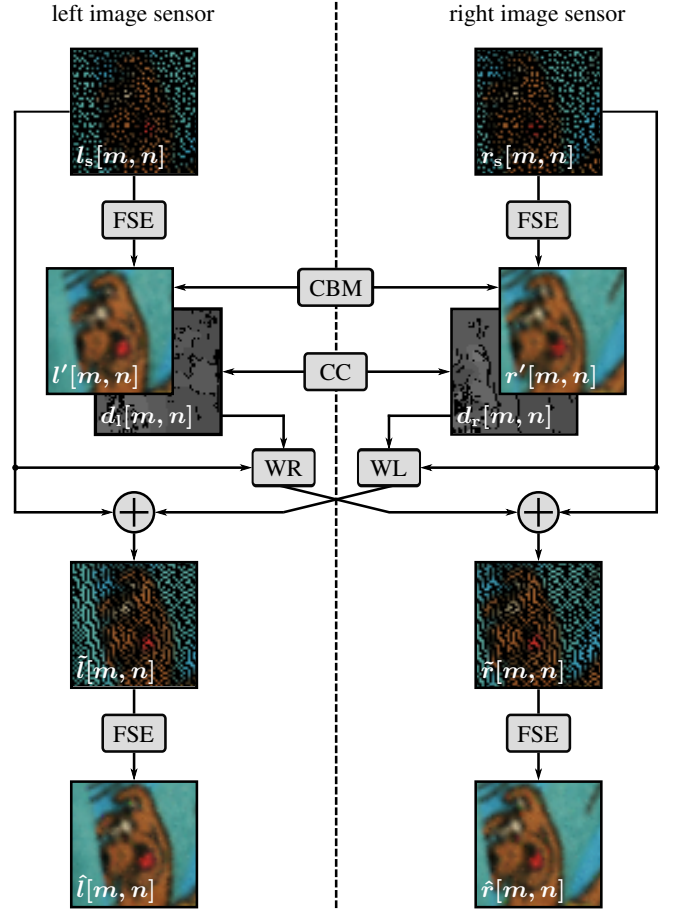
	\caption{Proposed stereo image reconstruction setup.}
	\label{fig:stereo_reconstruction_setup}
\end{figure}
In Fig.~\ref{fig:stereo_reconstruction_setup}, an overview of the proposed stereo reconstruction setup is given. A scene is captured by a pair of non-regular sampling sensors with low resolution yielding the sampled views $l_{\mathrm{s}}[m,n]$ for the left image sensor and $r_{\mathrm{s}}[m,n]$ for the right image sensor. A first version of fully reconstructed images $l^{\prime}[m,n]$ and $r^{\prime}[m,n]$ is obtained after FSE is applied independently to the sampled views. Up to this step, this corresponds to a single-view reconstruction 
and is from now on referred to as FSE-SV.

For projecting pixels from one image into the other image, disparity maps are required that are computed using CBM. Due to perspective distortions, textureless regions, and other challenges, matching is a difficult task and a lot of algorithms are available~\cite{Hirschmueller2007} to cope with these drawbacks. To obtain the disparity maps, both the left and the right image are scanned for matching features. A small square window $\mathcal{W}$ around the pixel of interest in one image is taken to find the corresponding pixel in the other image within a certain disparity range $\mathbf{d} = (d_{\mathrm{x}},d_{\mathrm{y}})$. Sum of absolute differences
\begin{equation}
	\text{SAD} = \sum_{x,y\in\mathcal{W}}{\left| l\left[x,y\right]  - r\left[x+d_{\mathrm{x}},y+d_{\mathrm{y}}\right] \right|}
\end{equation}
is one possible matching cost function for measuring the similarity between these windows and is used in the proposed stereo reconstruction setup for evaluating the disparities. $(x,y)$ depict the pixel within the window $\mathcal{W}$. The disparity associated with the lowest SAD value is considered the best match for corresponding features in both images. Scanning is done in both directions resulting in two disparity maps $\mathbf{d}_{\mathrm{l}}[m,n]$ and $\mathbf{d}_{\mathrm{r}}[m,n]$ which are computed with integer precision. For refining the disparity maps, a left-right consistency check (CC)~\cite{Hirschmueller2008} is performed to avoid half-occluded pixels in the final disparity maps. Half-occluded means that there is an object in one image and not in the other. To detect half-occluded pixels, a computed disparity value from one image is re-projected into the other image and compared to the corresponding disparity value from the other image. If both values differ, the pixel is considered half-occluded and the disparity is therefore set to invalid.

Employing forward-warping, the sampled view from one image sensor is warped to the adjacent sensor. Since the disparity maps hold the information how a pixel has to be shifted, disparity maps are used to obtain synthesized images for the opposite view. The warping is done from left to right (WR) and vice versa (WL). Where valid pixel information from the warped sampled views is available, it is used to fill the missing area in the original sampled views. This results in the merged sampled views $\tilde{l}[m,n]$ and $\tilde{r}[m,n]$. These images include now an extended set of sampling points which can be used as additional support during the last reconstruction step where FSE gives the final reconstructed images $\hat{l}[m,n]$ and $\hat{r}[m,n]$.

\begin{figure}[!t]
	\centering
	\input{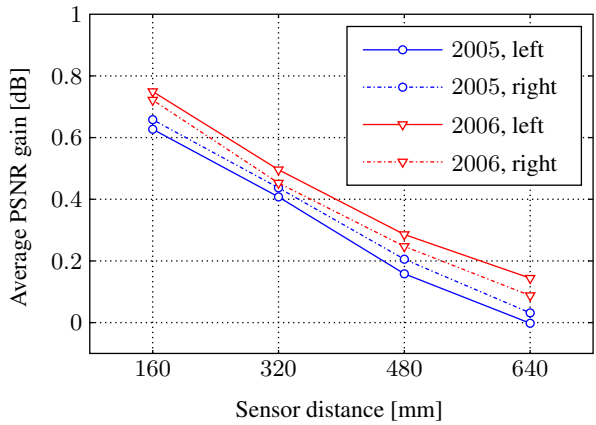}
	\vspace*{-0.1cm}
	\caption{Comparison of the proposed stereo image reconstruction against FSE-SV showing the average gain in PSNR for both views over the sensor distance and for Middlebury datasets $2005$ and $2006$.}
	\label{fig:plot_results}
	\vspace*{-0.3cm}
\end{figure}

\begin{figure*}
	\centering
	\def\svgwidth{\textwidth}
	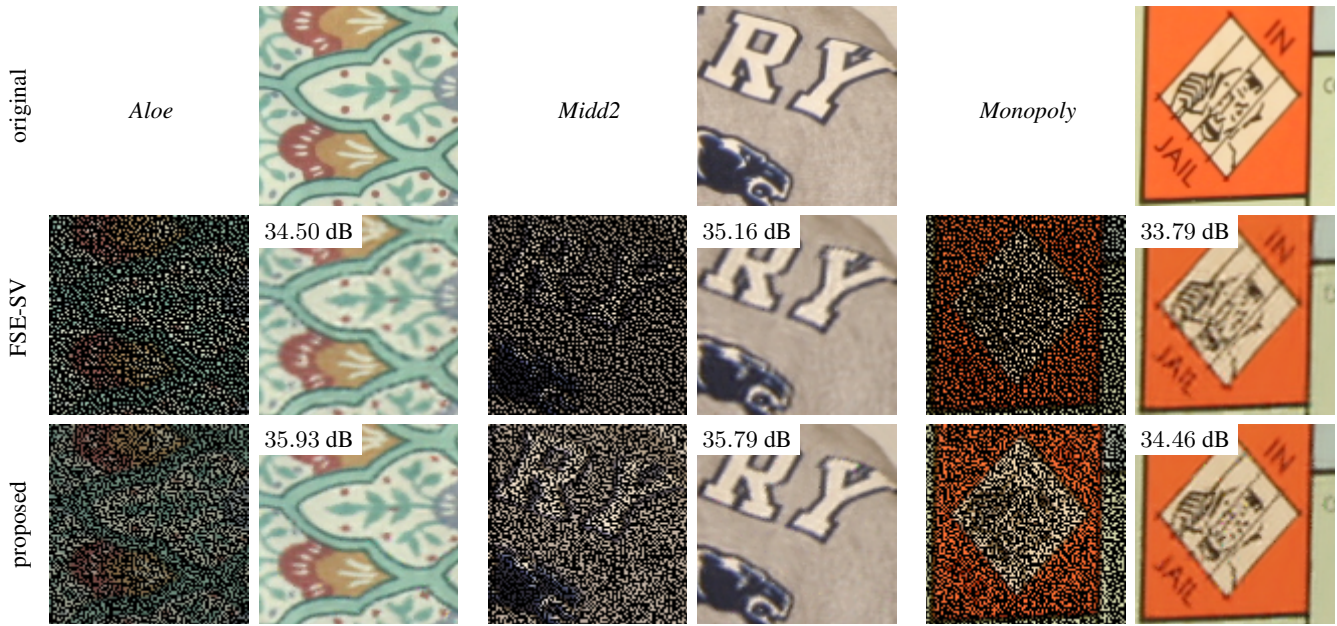
	\vspace*{-0.6cm}
	\caption{Three image detail examples comparing the single-view FSE approach (FSE-SV) and the proposed stereo reconstruction.}
	\label{fig:image_examples}
	\vspace*{-0.3cm}
\end{figure*}

\vspace*{-0.2cm}
\section{Simulation Results}
\label{sec:results}
\vspace*{-0.2cm}
The proposed stereo reconstruction setup has been tested for the Middlebury stereo dataset~\cite{Hirschmueller2007}. It contains $21$ images from the $2006$ dataset and $9$ images from the $2005$ dataset. For each image there exist $7$ views $(0,\ldots,6)$ taken under three different illuminations \mbox{({\it{Illum}})} and with three different exposure times ({\it{Exp}}). During simulation, only {\it{Illum$2$}} and {\it{Exp$1$}} have been used and all images have been converted to grayscale. The baseline of the camera and therefore the distance between the image sensors is $160$ mm. For the left image sensor {\it{view$1$}} has been used and for the right image sensor {\it{view$2$ - view$5$}}, depending on the sensor distance.

Following the reconstruction setup of Fig.~\ref{fig:stereo_reconstruction_setup}, sampled images for both views are generated using the same non-regular sampling mask on the original HR images which are only theoretically available. After the first reconstruction step, CBM is applied. Synthesized views are then computed from the sampled views and the disparity maps. Merging the missing parts of the sampled views with valid pixel information from the synthesized views leads to new sampled images. These images contain more reliable pixel information that can be used in the final reconstruction step. 
During simulation, the following parameters for FSE are used: $4 \times 4$ blocks to be reconstructed centered within blocks of size $28 \times 28$. The weighting function decays with $0.7$. Previously reconstructed pixels are weighted by $0.5$ and the Orthogonality Deficiency Compensation factor is set to $0.5$. A maximum number of $100$ iterations is chosen and the size of FFT basis functions is set to $32$. 

In Fig.~\ref{fig:plot_results}, overall results for both views and for the two different datasets are illustrated. The gain in PSNR as an average for the whole dataset is plotted over the sensor distance. It can be seen that the proposed stereo reconstruction setup gives similar results for both datasets and both views and has the largest gain for a sensor distance of $160$ mm. It is also noticeable that the larger the distance of the image sensors, the lower the overall gain gets.
In Table~\ref{tab:psnr_results}, PSNR results for several images from the $2006$ dataset are depicted. All values are taken for the left image sensor and for a sensor distance of $160$ mm. There is an average gain of $0.74$~dB for the whole $2006$ dataset and up to $1.43$ dB gain are possible for a {\it Aloe}. However, for {\it Plastic}, small negative gains can be observed. These losses occur only when \mbox{FSE-SV} already gives a very good reconstruction quality. Despite the negative values for the gain, these reconstructed images still have PSNR values of more than $47$ dB and show almost no differences in visual quality.
In addition to PSNR, Structural Similarity Index (SSIM)~\cite{Wang2004} has been used as a subjective measure. The resulting SSIM index lies between $0$ and $1$, whereas $1$ is only reachable in case of two identical images. Beside PSNR results, Table~\ref{tab:psnr_results} shows  the corresponding SSIM values. They have almost the same behavior as PSNR, however for images where there is a loss in PSNR, SSIM gains are still positive or equal to zero. According to SSIM, there is at least no loss in terms of visual quality.

Fig.~\ref{fig:image_examples} shows three image detail examples for the left image sensor comparing \mbox{FSE-SV} and the proposed stereo reconstruction setup. The first row contains the original images which are only theoretically available. The second row displays the images reconstructed by \mbox{FSE-SV} and the last row the results for the proposed reconstruction setup. For each approach, the corresponding sampling masks are shown and it can be clearly seen that in case of the proposed method, more valid pixel information is available that can be used in the final reconstruction step. Regarding \mbox{FSE-SV}, it is noticeable that for all images some fine details cannot be reconstructed and ringing artifacts or blurring may occur. With the proposed image reconstruction, all images appear sharper, fine details are reconstructed and especially small text is better readable.

Since FSE is computationally rather expensive, a linear interpolation technique can also be used in the first reconstruction step. This speedup, however, is paid with a loss in both PSNR and visual quality regarding the final reconstructed images. Compared to using FSE in both reconstruction steps, employing linear interpolation leads to an average loss in PSNR of about $0.2$ dB.

Using the proposed stereo image reconstruction setup, a visually noticeable gain in PSNR can be reached. Since FSE reconstruction highly depends on the number of available sampling points, it is expected that the better the estimation of the disparity maps, the better the final reconstruction quality gets. This could be achieved either by using a more complex correlation based matching method
or by additional refinements on the disparity maps. It can also be stated that the better the first reconstruction, the better the estimation of the disparity maps gets. Therefore, FSE should be applied for all reconstruction steps.

\vspace*{-0.2cm}
\section{Conclusion}
\label{sec:conclusion}
\vspace*{-0.2cm}
In this paper, a stereo image reconstruction setup has been proposed that captures a scene by a pair of non-regular sampling sensors. The idea of non-regular sampling provides the possibility to reduce costs in such a stereo scenario by avoiding the need for high resolution sensors. Using valid pixel information from neighboring views results in a better image reconstruction quality. Compared to a state-of-the-art single-view reconstruction algorithm, this leads to a visually noticeable average gain in PSNR of $0.74$ dB.

\newpage
\bibliographystyle{IEEEbib}
\bibliography{literature}

\begin{thebibliography}{1}

\bibitem{Seitz2006}
Steven~M. Seitz, Brian Curless, James Diebel, Daniel Scharstein, and Richard
  Szeliski,
\newblock ``{A} {C}omparison and {E}valuation of {M}ulti-{V}iew {S}tereo
  {R}econstruction {A}lgorithms,''
\newblock in {\em Proc. IEEE Conference on Computer Vision and Pattern
  Recognition}, New York, NY, USA, June 2006, vol.~1, pp. 519--528.

\bibitem{Dallmeier2012}
Dieter Dallmeier,
\newblock ``{S}urveillance apparatus using a plurality of image sensors,''
  December 2012,
\newblock CA Patent App. CA 2,805,079.

\bibitem{Gere2013}
David~S. Gere,
\newblock ``{I}mage capture using luminance and chrominance sensors,'' July
  2013,
\newblock US Patent 8,497,897.

\bibitem{Schoeberl2011a}
Michael Schöberl, Jürgen Seiler, Siegfried Foessel, and André Kaup,
\newblock ``{I}ncreasing {I}maging {R}esolution by {C}overing {Y}our
  {S}ensor,''
\newblock in {\em Proc. IEEE International Conference on Image Processing},
  Brussels, Belgium, September 2011, pp. 1937--1940.

\bibitem{Seiler2010}
Jürgen Seiler and André Kaup,
\newblock ``{C}omplex-{V}alued {F}requency {S}elective {E}xtrapolation for
  {F}ast {I}mage and {V}ideo {S}ignal {E}xtrapolation,''
\newblock {\em Signal Processing Letters}, vol. 17, no. 11, pp. 949--952,
  November 2010.

\bibitem{Seiler2011}
Jürgen Seiler and André Kaup,
\newblock ``{O}ptimized and {P}arallelized {P}rocessing {O}rder for {I}mproved
  {F}requency {S}elective {S}ignal {E}xtrapolation,''
\newblock in {\em Proc. European Signal Processing Conference}, Barcelona,
  Spain, August 2011, pp. 269--273.

\bibitem{Hirschmueller2007}
Heiko Hirschmüller and Daniel Scharstein,
\newblock ``{E}valuation of {C}ost {F}unctions for {S}tereo {M}atching,''
\newblock in {\em Proc. IEEE Conference on Computer Vision and Pattern
  Recognition}, Minneapolis, MN, USA, 2007, pp. 1--8.

\bibitem{Hirschmueller2008}
Heiko Hirschmüller,
\newblock ``{S}tereo {P}rocessing by {S}emiglobal {M}atching and {M}utual
  {I}nformation,''
\newblock {\em IEEE Transactions on Pattern Analysis and Machine Intelligence},
  vol. 30, no. 2, pp. 328--341, February 2008.

\bibitem{Wang2004}
Zhou Wang, A.C. Bovik, H.R. Sheikh, and E.P. Simoncelli,
\newblock ``{I}mage {Q}uality {A}ssessment: {F}rom {E}rror {V}isibility to
  {S}tructural {S}imilarity,''
\newblock {\em IEEE Transactions on Image Processing}, vol. 13, no. 4, pp.
  600--612, April 2004.

\end{thebibliography}
\label{sec:ref}

\end{document}